\input phyzzx
\vsize=23.5cm
\singlespace
\vsize=23.5cm
\baselineskip14.6pt

\centerline{\bf Recent Progress in Nonperturbative QCD Theory} 

\centerline{\bf and Insight on Cosmological Phase Transition}

\vskip0.5cm

\centerline{H.~Suganuma, H.~Ichie, H.~Toki and H.~Monden$^*$}

\vskip0.5cm

\centerline{\it Research Center for Nuclear Physics (RCNP), 
Osaka University}
\centerline{\it Mihogaoka 10-1, Ibaraki 567, Osaka, Japan}

\vskip0.25cm

\centerline{\it $^*$ Department of Physics, 
Tokyo Metropolitan University}
\centerline{\it Minami-Osawa 1-1, Hachioji 192, Tokyo, Japan} 

\vskip0.5cm

\tenpoint

The QCD phase transition at finite temperature 
is studied with the dual Ginzburg-Landau theory, 
which is the QCD effective theory based on  
the dual Higgs mechanism by QCD-monopole condensation.
At high temperature, the confinement force 
is largely reduced by thermal effects, 
which leads to the swelling of hadrons. 
Simple formulae for the surface tension and the 
thickness of the phase boundary are derived from 
the shape of the effective potential at the critical temperature.
We investigate also the process of the hadron-bubble formation 
in the early Universe.

\twelvepoint
\twelverm

\chapter{Introduction}

Quantum chromodynamics (QCD) is the fundamental theory 
of the strong interaction 
\REF\greiner{
W.~Greiner and A.~Sch\"afer, ``Quantum Chromodynamics'', 
(Springer-Verlag, 1994).
}
\REF\cheng{
T.~P.~Cheng and L.~F.~Li, 
``Gauge Theory of Elementary Particle Physics"
(Clarendon press, Oxford, 1984).
}
\REF\huang{
K.~Huang, ``Quarks, Leptons and Gauge Fields'', 
(World Scientific, 1982).
}
[\greiner-\huang].
In spite of the simple form of the QCD lagrangian,
$$
{\cal L}_{\rm QCD}=-{1 \over 2} {\rm tr} G_{\mu \nu }G^{\mu \nu }
+\bar q (\not D-m_q) q, 
\eqn\qcd
$$
it miraculously provides quite various phenomena 
like color confinement, dynamical chiral-symmetry breaking, 
non-trivial topologies, quantum anomalies and so on, as shown 
in Fig.1.
Then, it would be interesting to compare QCD with the history of 
the Universe, because a quite simple `big bang' 
also created various things including galaxies, stars, lives and 
thinking reeds 
\REF\pascal{
B.~Pascal, ``Pens\'ees'' (1670).
}
[\pascal]. 
Therefore, QCD can be regarded as an interesting 
miniature of the history of the Universe. 
This is the most attractive point of the QCD physics.

Since it is quite difficult to understand the various QCD phenomena and 
their underlying mechanism at the same time, many methods 
and models have been proposed to understand each phenomenon.
We show in Fig.2 a brief sketch on the history of QCD and 
typical QCD effective models [\greiner] as listed in Table 1. 
The bag model and the Nambu-Jona-Lasinio (NJL) 
model are simple relativistic quark theories. 
The nonlinear $\sigma$ model and the Skyrme-Witten 
soliton model are described by the Nambu-Goldstone pion. 
In particular, the Skyrme-Witten model is stimulative because a 
fermionic baryon can be described as a soliton made by bosonic pions 
in this framework. The origin of this magic is found in 
the non-trivial topology of the flavor dynamics in QCD. 
The lattice QCD simulation 
\REF\rothe{
H.~J.~Rothe, ``Lattice Gauge Theories'', (World Scientific, 1992).
}
[\rothe] 
using a supercomputer is a hopeful and promising method based on 
QCD directly, and its importance becomes larger and larger 
according to the great progress of the computational power. 
In particular, recent lattice QCD studies 
\REF\hioki{
A.~S.~Kronfeld, G.~Schierholz and U.~-J.~Wiese, 
Nucl.~Phys.~{\bf B293} (1987) 461.
\nextline
S.~Hioki, S.~Kitahara, S.~Kiura, Y.~Matsubara, 
O.~Miyamura, S.~Ohno and 
\nextline
T.~Suzuki, Phys.~Lett.~{\bf B272} (1991) 326. 
\nextline
S.~Kitahara, Y.~Matsubara and T.~Suzuki, 
Prog.~Theor.~Phys.~{\bf 93} (1995) 1.
}
[\hioki] 
shed a light on the confinement mechanism, which 
is one of the most difficult problems in the particle physics. 
In these years, the origin of color confinement can be recognized 
as the dual Higgs mechanism by monopole condensation, 
and the nonperturbative QCD vacuum is regarded as 
the dual superconductor.
The dual Ginzburg-Landau (DGL) theory 
\REF\suzuki{
T.~Suzuki, Prog.~Theor.~Phys. {\bf 80} (1988) 929 ; {\bf 81} (1989) 752.
\nextline
S.~Maedan and T.~Suzuki, Prog.~Theor.~Phys. {\bf 81} (1989) 229. 
}
\REF\suganumaA{
H.~Suganuma, S.~Sasaki and H.~Toki, Nucl.~Phys.~{\bf B435} (1995) 207. 
}
\REF\suganumaB{
H.~Suganuma, S.~Sasaki and H.~Toki, 
Proc. of Int.~Conf. on 
``Quark Confinement and Hadron Spectrum", Como Italy, June 1994,
(World Scientific, 1995) 238.
}
\REF\suganumaC{
H.~Suganuma, S.~Sasaki, H.~Toki and H.~Ichie, 
Prog.~Theor.~Phys.~(Suppl.)~{\bf 120} (1995) 57.  
}
\REF\suganumaD{
H.~Suganuma, H.~Ichie, S.~Sasaki and H.~Toki, 
``Color Confinement and Hadrons", 
(World Scientific, 1995) 65.
}
\REF\ichieA{
H.~Ichie, H.~Suganuma and H.~Toki, 
Phys.~Rev.~{\bf D52} (1995) 2944.
}
\REF\suganumaE{
H.~Suganuma, S.~Umisedo, S.~Sasaki, H.~Toki and O.~Miyamura,
Proc. of ``Quarks, Hadrons and Nuclei", 
Adelaide Australia, Nov. 1995, to appaer in Aust.~J.~Phys. 
}
\REF\ichieB{
H.~Ichie, H.~Monden, H.~Suganuma and H.~Toki, 
Proc. of ``Nuclear Reaction Dynamics of Nucleon-Hadron Many Body 
Dynamics", Osaka, Dec. 1995 (World Scientific) in press.
}
[\suzuki-\ichieB]
was formulated with this picture, and can reproduce 
confinement properties like the string tension 
and the hadron flux-tube formation.

Now, you may find a current of the QCD physics. 
In '80s, chiral symmetry breaking was the central issue. 
The chiral bag model, the NJL model and the $\sigma$ model were 
formulated with referring chiral symmetry.
In '90s, on the other hand, the confinement physics is providing 
an important current of the hadron physics. 
The key word for the understanding of confinement 
is the ``duality'', which is recently paid attention by 
many theoretical particle physicists after Seiberg-Witten's discovery 
on the essential role of monopole condensation for the confinement 
in a supersymmetric version of QCD 
\REF\seiberg{
N.~Seiberg and E.~Witten, Nucl.~Phys.~{\bf B426} (1994) 19; 
{\bf B431} (1994) 484. 
\nextline
G.~P.~Collins, Physics Today, March (1995) 17.
}
[\seiberg].

\chapter{Color Confinement and Dual Higgs Mechanism}

We briefly show the modern current of the confinement physics. 
About 20 years ago, Nambu, 't~Hooft and Mandelstam proposed an 
interesting picture for color confinement based on the analogy 
between the superconductor and the QCD vacuum 
\REF\nambu{
Y.~Nambu, Phys.~Rev.~{\bf D10} (1974) 4262.
\nextline
G.~'t~Hooft, ``High Energy Physics" 
(Editorice Compositori, Bologna 1975).
\nextline
S.~Mandelstam, Phys.~Rep.~{\bf C23} (1976) 245.
}
[\nambu].
In the superconductor, the magnetic field is excluded 
due to the Meissner effect, which is caused by Cooper-pair condensation.
As the result, the magnetic flux is squeezed like the Abrikosov vortex.
On the other hand, the color-electric flux is excluded 
in the QCD vacuum, and therefore the squeezed color-flux tube is 
formed between color sources. Thus, these two systems are quite 
similar and can be regarded as the dual version each other.
This idea is based on the ``duality'' of the gauge theories, which was 
firstly pointed out by Dirac more than 50 years ago 
\REF\dirac{
P.~A.~M.~Dirac, Proc.~Roy.~Soc.~{\bf A133} (1931) 60. 
}
[\dirac].

With referring Table 2 and Fig.3, we compare the ordinary 
electromagnetic system, the superconductor and 
the nonperturbative QCD vacuum regarded as the dual superconductor. 
In the ordinary electromagnetism in the Coulomb phase, 
both electric flux and magnetic flux 
are conserved, respectively. The electric-flux conservation 
is guarantied by the ordinary gauge symmetry. 
On the other hand, the magnetic-flux conservation 
is originated from the dual gauge symmetry [\suganumaA-\ichieB], 
which is the generalized version of the Bianchi identity. 
As for the inter-charge potential in the Coulomb phase, 
both electric and magnetic potentials are Coulomb-type. 

The superconductor in the Higgs phase is characterized by 
electric-charge condensation, which leads to the Higgs mechanism or 
spontaneous breaking of the ordinary gauge symmetry, 
and therefore the electric flux is no more conserved. 
In such a system obeying the London equation, 
the electric inter-charge potential becomes short-range Yukawa-type 
similarly in the electro-weak unified theory. 
On the other hand, the dual gauge symmetry is not broken, so that the 
magnetic flux is conserved, but is squeezed like a one-dimensional 
flux tube due to the Meissner effect.
As the result, the magnetic inter-charge potential becomes 
linearly rising like a condenser.

The nonperturbative QCD vacuum regarded as the dual Higgs phase 
is characterized by color-magnetic monopole condensation, 
and resembles the dual version of the superconductor, 
where the ``dual version'' means the interchange between 
the electric and magnetic sectors.
Monopole condensation leads to the spontaneous breaking of the 
dual gauge symmetry, so that color-magnetic flux is not conserved, 
and the magnetic inter-change potential becomes 
short-range Yukawa-type.
Note that the ordinary gauge symmetry is not broken by such monopole 
condensation. Therefore, color-electric flux is conserved, 
but is squeezed like a one-dimensional flux-tube or a string 
as a result of the dual Meissner effect.
Thus, the hadron flux-tube is formed in the monopole-condensed 
QCD vacuum, and the electric inter-charge potential 
becomes linearly rising, which confines the color-electric charges as 
quarks [\suganumaA-\suganumaD].

As a remarkable fact in the duality physics, 
these are two ``see-saw relations'' between in the electric and 
magnetic sectors. 
\nextline
(1) There appears the Dirac condition $eg=2\pi$ [\dirac] in QED 
or $eg=4\pi$ [\suganumaA] in QCD. 
Here, unit electric charge $e$ is the gauge coupling constant, 
and unit magnetic charge $g$ is the dual gauge coupling constant. 
Therefore, a strong-coupling system in one sector corresponds to 
a weak-coupling system in the other sector.
\nextline
(2) As shown in Fig.3, the long-range confinement system in one sector 
corresponds to a short-range (Yukawa-type) interaction system in the 
other sector.

Let us consider usefulness of the latter ``see-saw relation''.
One faces highly non-local properties among the color-electric 
charges in the QCD vacuum because of the long-range linear 
confinement potential.
Then, the direct formulation among the electric-charged variables 
would be difficult due to the non-locality, 
which seems to be a destiny in the long-distance confinement physics. 
However, one finds a short-range Yukawa potential 
in the magnetic sector, so that the electric-confinement 
system can be approximated by a local formulation 
among magnetic-charged variables. 
Thus, the confinement system, which seems highly non-local,  
can be described by a short-range interaction theory using 
the dual variables, which is the most attractive point 
in the dual Higgs theory.

Color-magnetic monopole condensation is necessary for 
color confinement in the dual Higgs theory. 
As for the appearance of color-magnetic monopoles in QCD, 't~Hooft 
proposed an interesting idea of the abelian gauge fixing
\REF\thooft{
G.~'t~Hooft, Nucl.~Phys.~{\bf B190} (1981) 455.
}
\REF\ezawa{
Z.~F.~Ezawa and A.~Iwazaki, Phys.~Rev.~{\bf D25} (1982) 2681; 
{\bf D26} (1982) 631.
}
[\thooft,\ezawa], which is defined by the diagonalization 
of a gauge-dependent variable. 
In this gauge, QCD is reduced into an abelian gauge theory with 
the color-magnetic monopole [\suganumaA,\thooft], 
which will be called as QCD-monopoles hereafter. 
Similar to the 't~Hooft-Polyakov monopole 
[\cheng] in the Grand Unified theory 
(GUT), the QCD-monopole appears from a hedgehog configuration 
corresponding to the non-trivial homotopy group 
$\pi_2({\rm SU}(N_c)/{\rm U}(1)^{N_c-1})=Z_\infty ^{N_c-1}$ 
on the nonabelian manifold. 

Many recent studies based on the lattice gauge theory have supported 
QCD-monopole condensation and abelian dominance 
\REF\miyamura{
O.~Miyamura, Nucl.~Phys.~{\bf B}(Proc.~Suppl.){\bf 42} (1995) 538.
\nextline
O.~Miyamura and S.~Origuchi, ``Color Confinement and Hadrons'', 
(World Scientific, 1995) 235.
}
\REF\suganumaF{
H.~Suganuma, A.~Tanaka, S.~Sasaki and O.~Miyamura, 
Proc. of ``Lattice Field Theories '95", 
Nucl.~Phys.~{\bf B}~(Proc.~Suppl.)~{\bf 47} (1996) 302.
}
\REF\suganumaG{
H.~Suganuma, K.~Itakura, H.~Toki and O.~Miyamura, 
Proc. of ``Non-perturbative Approaches to QCD", 
Trento Italy, July 1995 (PNPI, 1995) 224. 
}
[\hioki,\suganumaE,\miyamura-\suganumaG], 
which means that only abelian variables are relevant for 
the nonperturbative QCD, in the 't~Hooft abelian gauge. 
Hence, the dual-superconductor scenario seems 
workable for color confinement in the QCD vacuum, and 
the nonperturbative QCD would be described by the dual Ginzburg-Landau 
(DGL) theory, which is the QCD effective theory based on the 
dual Higgs mechanism.

\chapter{Dual Ginzburg-Landau Theory}

The dual Ginzburg-Landau (DGL) lagrangian 
[\suganumaE,\ichieB] 
for the pure-gauge system is described with 
the dual gauge field $B_\mu$ and the QCD-monopole field $\chi$, 
$$
{\cal L}_{\rm DGL}={\rm tr} \left\{
-{1 \over 2}(\partial _\mu  B_\nu -\partial _\nu B_\mu )^2
+[\hat{D}_\mu, \chi]^{\dag}[\hat{D}^\mu, \chi]
-\lambda ( \chi^{\dag} \chi -v^2)^2 \right\},
\eqn\TWO
$$
where $\hat{D}_\mu \equiv \hat{\partial}_\mu +i g B_\mu$ 
is the dual covariant derivative 
including the dual gauge coupling constant $g=4\pi/e$.

The dual gauge field 
$B_\mu \equiv \vec B_\mu \cdot \vec H = B_\mu^3 T^3+B_\mu^8 T^8$ 
is defined on the dual gauge manifold 
${\rm U(1)}_m^3 \times {\rm U(1)}_m^8$ [\suganumaE,\ichieB], 
which is the dual space of the maximal torus subgroup
${\rm U(1)}_e^3 \times {\rm U(1)}_e^8$ 
embedded in the original gauge group ${\rm SU(3)}_c$.
The abelian field strength tensor is written as  
$F_{\mu \nu}={}^* (\partial\wedge B)_{\mu\nu}$ 
so that the role of the electric and magnetic fields are interchanged 
in comparison with the ordinary $A_\mu$ description.

The QCD-monopole field $\chi$ is defined as 
$\chi \equiv \sqrt{2} \sum_{\alpha=1}^3 \chi_\alpha E_\alpha$ 
[\suganumaE,\ichieB] where 
$E_1 \equiv {1 \over {\sqrt{2}}}(T_6+iT_7)$,
$E_2 \equiv {1 \over {\sqrt{2}}}(T_4-iT_5)$ and 
$E_3 \equiv {1 \over {\sqrt{2}}}(T_1+iT_2)$.
Here, $\chi_\alpha$ has the magnetic charge $g\vec \alpha$ 
proportional to the root vector $\vec \alpha$. 
In the QCD-monopole condensed vacuum with $|\chi_\alpha| =v$, 
the dual gauge symmetry ${\rm U(1)}_m^3 \times {\rm U(1)}_m^8$ 
is spontaneously broken instead of 
the gauge symmetry ${\rm U(1)}_e^3 \times {\rm U(1)}_e^8$. 
Through the dual Higgs mechanism, 
the dual gauge field $B_\mu$ acquires its mass $m_B = \sqrt{3}g v $, 
whose inverse provides the radius of the hadron flux tube [\suganumaA], 
and the dual Meissner effect causes the color-electric field 
excluded from the QCD vacuum, which leads to color confinement. 
The QCD-monopole fluctuations 
$\tilde \chi_\alpha \equiv \chi_\alpha - v$ ($\alpha$=1,2,3) 
also acquire their mass $m_\chi= 2 \sqrt{\lambda} v$ in the 
QCD-monopole condensed vacuum. 
As a relevant prediction, only one QCD-monopole fluctuation 
$\tilde \chi_ \equiv \sum_{\alpha=1}^3 \tilde \chi_\alpha$ appears as a 
color-singlet scalar glueball in the confinement phase, 
although the dual gauge field $B_\mu$ and 
the other two combinations of the QCD-monopole fluctuation 
are not color-singlet and cannot be observed 
[\suganumaE,\suganumaF,\suganumaG]. 

The DGL theory reproduces confinement properties like 
the inter-quark potential and the hadron flux-tube formation. 
We studied effects of the flux-tube breaking 
by the light quark-pair creation in the DGL theory, 
and derived the infrared screened inter-quark potential 
[\suganumaA-\suganumaC], 
which is observed in the lattice QCD with dynamical quarks [\rothe]. 
We studied also the dynamical chiral-symmetry breaking (D$\chi$SB), 
which is also an important feature in the nonperturbative QCD, 
by solving the Schwinger-Dyson equation for the dynamical quark 
[\suganumaA-\suganumaC]. 
The quark-mass generation is brought by QCD-monopole condensation, 
which suggests the close relation between D$\chi$SB 
and color confinement.
Thus, the DGL theory provides not only the confinement properties 
but also D$\chi$SB and its related quantities like the constituent 
quark mass, the chiral condensate and the pion decay constant 
\REF\sasakiA{
S.~Sasaki, H.~Suganuma and H.~Toki, Prog.~Theor.~Phys. {\bf 94} (1995) 
373.
}
\REF\sasakiB{
S.~Sasaki, H.~Suganuma and H.~Toki, Proc. of Int.~Conf.~on 
``Baryons '95", Santa Fe, Oct. 1995, in press.
}
[\suganumaA-\suganumaC,\sasakiA,\sasakiB].

\chapter{QCD Phase Transition in the DGL Theory}

In this chapter, we study the QCD phase transition 
[\suganumaC-\ichieA,\ichieB,\sasakiB] in the DGL theory.
Although quarks and gluons are confined inside hadrons in the 
nonperturbative QCD vacuum, 
these colored particles are liberated 
and the system becomes the quark-gluon-plasma (QGP) phase 
[\greiner]
at high temperature ($T > T_c \sim 200{\rm MeV}$), 
which is called as the QCD phase transition. 
The experimental creation of the QGP, 
one of the most central subjects in the RHIC project, 
is expected to be realized in ultra-relativistic heavy-ion 
collisions 
\REF\ichieC{
H.~Ichie, H.~Suganuma and H.~Toki, hep-ph/9602412, 
Phys.~Rev.~{\bf D} in press. 
}
[\greiner,\ichieC].
On the other hand, the QCD phase transition occurred 
in the early Universe, and its process strongly influenced 
the afterward nucleosynthesis 
\REF\kajinoA{
T.~Kajino and R.~N.~Boyd, Ap.~J.~{\bf 359} (1990) 267.
\nextline
T.~Kajino, G.~J.~Mathew and G.~M.~Fuller, Ap.~J.~{\bf 364} (1990) 7.
}
[\kajinoA].

In the DGL theory, the QCD phase transition is characterized by 
QCD-monopole condensate $\bar \chi \equiv |\chi_\alpha|$,
which is an order parameter on the confinement strength. 
The QCD-monopole condensed vacuum ($\chi \ne 0$), where the dual 
gauge symmetry is spontaneously broken ($m_B \ne 0$), is the 
confinement phase with a non-vanishing string tension ($k \ne 0$). 
Without QCD-monopole condensation ($\chi = 0$), 
the dual gauge symmetry is manifest ($m_B=0$), and the system 
corresponds to the deconfinement phase, where the 
string tension disappears ($k=0$). 

To study the QCD phase transition, the effective potential 
$V_{\rm eff}(\bar \chi;T)$ at finite temperature, 
which physically means the thermodynamical potential, 
is formulated as the function of QCD-monopole condensate 
$\bar \chi \equiv |\chi_\alpha|$ [\suganumaC-\ichieA,\ichieB], 
$$
V_{\rm eff}(\bar \chi;T) = 3 \lambda ( \bar \chi^2 - v^2 )^2 
            + {3T} \int {d^3k \over (2 \pi)^3} 
            [2 \ln (1 - e^{ - \sqrt{ k^2 + m_B^2}/T})
            + \ln (1 - e^{ - \sqrt{ k^2 + m_{\chi}^2}/T })]
\eqn\THREE
$$
with $m_B= \sqrt{3} g \bar \chi$ and $m_\chi=2\sqrt{\lambda} 
\bar \chi$. Here, we have used the quadratic source term 
to avoid the imaginary scalar-mass problem [\suganumaC-\ichieA,\ichieB]. 

The QCD-monopole condensate $\bar \chi_{\rm phys}(T)$ 
at finite temperature is obtained from the local 
minimum of the effective potential $V_{\rm eff}(\bar \chi;T)$. 
As temperature increases, $\bar \chi_{\rm phys}(T)$ decreases and 
disappears at a critical temperature $T_c$.
Above $T_c$, the deconfinement phase is realized as 
the Coulomb phase with $\bar \chi_{\rm phys}(T)=0$, where 
the dual gauge symmetry is restored.
With the parameters in Ref.[\suganumaA], 
we find a weak first order phase transition at $T_c$=0.2GeV, and  
the mixed phase or the two-phase coexistence is allowed only in 
$T_{\rm low}< T <T_{\rm up}$ with $T_{\rm low}\simeq$ 0.193GeV and 
$T_{\rm up}\simeq$ 0.201GeV [\suganumaC,\ichieA,\ichieB]. 

Now, let us consider the confinement and hadron properties.
Fig.4 shows the string tension $k(T)$ at finite temperature. 
As the temperature increases, $k(T)$ becomes smaller and drops 
rapidly near $T_c$ 
\REF\gao{
M.~Gao, Nucl.~Phys.~{\bf B9} (Proc. Suppl.) (1989) 368.
}
[\suganumaC-\ichieA,\ichieB,\gao].
Therefore, the slope of the inter-quark potential is reduced 
and the inter-quark distance inside hadrons 
increases at high temperature. 
In addition, the color-electric field spreads according 
to the decrease of $m_B$.
Thus, the reduction of the confinement force leads to the swelling of 
hadrons at high temperature [\ichieB].

We predict also a large mass reduction of the scalar 
glueball originated from QCD-monopoles near $T_c$ 
[\suganumaC-\ichieA,\ichieB].
We guess that the violent excitation of these scalar glueballs 
with a reduced mass would promote the QCD phase transition. 

Finally, we consider the relation between the surface tension $\sigma$ 
and the effective potential $V_{\rm eff}(\bar \chi;T_c)$ 
[\suganumaC,\suganumaD]. 
The surface tension $\sigma$ characterizes the strength of 
the first order in the phase transition, and is very important 
for the shape of the boundary surface in the mixed phase, where 
the two phases correspond to the two minima, $\bar \chi =0$ 
and $\bar \chi = \bar \chi_c$, in $V_{\rm eff}(\bar \chi;T_c)$. 
Using the sine-Gordon kink ansatz [\suganumaC,\suganumaD] 
for the boundary profile, 
$
\bar \chi (z)= \bar \chi_c \tan ^{-1}e^{z/\delta}, 
$
we derive simple formulae for the surface tension 
$
\sigma \simeq {4\sqrt{3} \over \pi} \sqrt{h} \bar \chi_c, 
$
and the phase-boundary thickness 
$
2\delta \simeq {2\sqrt{3} \over \pi} \bar \chi_c/\sqrt{h}, 
$
where $h$ is the ``barrier height" between the two minima in 
$V_{\rm eff}(\bar \chi;T_c)$. 
We find $2\delta \simeq 3.4 {\rm fm}$ and 
$\sigma \simeq (112{\rm MeV})^3 $ [\suganumaC], 
which seems consistent with the lattice QCD data 
\REF\kanaya{
Y.~Iwasaki, K.~Kanaya, L.~Karkkainen, Phys.~Rev.~{\bf D49} (1994) 3540.
}
[\kanaya].

\chapter{Hadron Bubble Formation in the Early Universe}

In this chapter, we study the hadron bubble formation 
\REF\fuller{
G.~M.~Fuller, G.~J.~Mathews, and C.~R.~Alcock,
Phys.~Rev.~{\bf D 37} (1988) 1380.
}
[\fuller] 
in the early Universe using the DGL theory [\ichieB].
As Witten pointed out 
\REF\witten{
E.~Witten, Phys.~Rev.~{\bf D30}~(1984) 272. 
}
[\witten], 
if the QCD phase transition is of the first order, the hadron and QGP 
phases should coexist in the early Universe. 
During the mixed-phase period, there appears 
the inhomogeneity on the baryon density distribution 
\REF\kajinoB{
T.~Kajino, Phys.~Rev.~Lett.~{\bf 66} (1991) 125.
\nextline
T.~Kajino, M.~Orito, Y.~Yamamoto and H.~Suganuma, 
``Color Confinement and Hadrons", (World Scientific, 1995) 263.
}
[\kajinoA,\kajinoB], 
which can strongly affects the primordial nucleo-synthesis 
and the successive history of the Universe.

Let us consider how hadron bubbles appear in the QGP phase 
near the critical temperature $T_c$ in the DGL theory. 
The hadron bubbles are created in the supercooling QGP phase with 
$T_{\rm low} < T < T_{\rm c}$. 
We use the sine-Gordon kink ansatz [\ichieB] for the profile of 
the QCD-monopole condensate in the hadron bubble 
as a function of the radial coordinate $r$, 
$
\bar \chi (r;R)= \bar \chi_{\rm phys}(T) 
\tan^{-1}e^{(R-r)/\delta}/ \tan^{-1}e^{R/\delta}, 
$
where $R$ and $2\delta$ correspond to the hadron-bubble radius 
and the phase-boundary thickness, respectively. 
The QCD-monopole condensate $\bar \chi(r;R)$ 
is finite only inside the bubble, $r \lsim R$. 
The energy density ${\cal E}(r;R)$ of the hadron bubble is shown 
in Fig.5. It is negative inside and positive near the boundary surface.
The total energy is roughly estimated as the sum of the positive 
surface term and the negative volume term.

The total energy of the hadron bubble with radius $R$ can be 
estimated using the effective potential $V_{\rm eff}(\bar \chi;T)$, 
$
E(R;T) = 4 \pi \int_0^{\infty}drr^2
\{ 3( {d\bar \chi(r;R) \over dr})^2 + V_{\rm eff}(\bar \chi ;T)\}, 
$
where the thickness $\delta$ is determined by the energy minimum 
condition.
The hadron-bubble energy $E(R;T)$ as shown in Fig.6 
takes a maximal value at a critical radius $R_c$, which leads 
the collapse of the hadron bubbles with smaller radius than $R_c$. 
Only large hadron bubbles with $R > R_c$ grow up with radiating 
the shock wave [\ichieB,\fuller]. 

On the other hand, the creation of large bubbles is suppressed 
because of the small creation probability [\ichieB]. 
In the hadron-bubble formation, 
there is a penetration over a large barrier height 
$h(T)$ in the effective potential per unit volume, and therefore 
the creation of large bubbles needs a large energy fluctuation.
Such a process is strongly suppressed 
because of the thermodynamical factor 
$P(T) \equiv {\rm exp}(-{4 \pi \over 3} R_{\rm c}(T)^3 h(T)/T )$, 
which is proportional to the hadron-bubble formation rate. 
Thus, the only small bubbles are created practically, 
although its radius should be larger than $R_{\rm c}$ [\ichieB].

Using $V_{\rm eff}(\bar \chi;T)$ in the DGL theory, 
we can estimate the critical radius $R_c(T)$ 
and the hadron-bubble formation factor $P(T)$ at finite temperature 
as shown in Fig.7 and Fig.8, respectively. 
As the temperature decreases, the created hadron bubbles 
becomes smaller, while the bubble formation rate becomes larger 
[\ichieB].

From these results, we can imagine how the QCD phase transition 
happens in the big bang scenario [\ichieB] as shown in Fig.9. 

\item{\rm (a)} Slightly below $T_{\rm c}$, 
only large hadron bubbles appear, but the creation rate is quite small.
\item{\rm (b)} As temperature is lowered by the expansion of the Universe, 
smaller bubbles are created with much formation rate.
During this process, the created hadron bubbles expand with 
radiating shock wave, which reheats the QGP phase around them [\fuller].
\item{\rm (c)} Near $T_{\rm low}$, many small hadron bubbles 
are violently created in the unaffected region free from the shock wave. 
\item{\rm (d)} The QGP phase pressured by the hadron phase 
is isolated as high-density QGP bubbles [\fuller,\witten], 
which provide the baryon density fluctuation [\kajinoA]. 

\noindent
Thus, the numerical simulation using the DGL theory would tell 
how the hadron bubbles appear and evolve 
quantitatively in the early Universe. 

\vskip0.5cm

We would like to thank Prof.~Kajino for fruitful discussions on 
the hadron bubble formation.

\vskip0.5cm

{\bf REFERENCES}

\vskip0.2cm

\item{\rm 1.}
W.~Greiner and A.~Sch\"afer, ``Quantum Chromodynamics'', 
(Springer, 1994).
\item{\rm 2.}
T.~P.~Cheng and L.~F.~Li, 
``Gauge Theory of Elementary Particle Physics"
(Clarendon press, Oxford, 1984).
\item{\rm 3.}
K.~Huang, ``Quarks, Leptons and Gauge Fields'', 
(World Scientific, 1982).
\item{\rm 4.}
B.~Pascal, ``Pens\'ees'' (1670).
\item{\rm 5.}
H.~J.~Rothe, ``Lattice Gauge Theories'', (World Scientific, 1992).
\item{\rm 6.}
A.~S.~Kronfeld, G.~Schierholz, U.~-J.~Wiese, 
Nucl.~Phys.~{\bf B293} (1987) 461.
\nextline
S.~Hioki, S.~Kitahara, S.~Kiura, Y.~Matsubara, 
O.~Miyamura, S.~Ohno and 
\nextline
T.~Suzuki, Phys.~Lett.~{\bf B272} (1991) 326. 
\nextline
S.~Kitahara, Y.~Matsubara and T.~Suzuki, 
Prog.~Theor.~Phys.~{\bf 93} (1995) 1.
\item{\rm 7.}
T.~Suzuki, Prog.~Theor.~Phys. {\bf 80} (1988) 929 ; {\bf 81} (1989) 752.
\nextline
S.~Maedan and T.~Suzuki, Prog.~Theor.~Phys. {\bf 81} (1989) 229. 
\item{\rm 8.}
H.~Suganuma, S.~Sasaki and H.~Toki, Nucl.~Phys.~{\bf B435} (1995) 207. 
\item{\rm 9.}
H.~Suganuma, S.~Sasaki and H.~Toki, 
Proc.~of~Int.~Conf.~on 
``Quark Confinement and Hadron Spectrum", Como Italy, 
(World Scientific, 1995) 238.
\item{\rm 10.}
H.~Suganuma, S.~Sasaki, H.~Toki and H.~Ichie, 
Prog.~Theor.~Phys.~(Suppl.) {\bf 120} (1995) 57.  
\item{\rm 11.}
H.~Suganuma, H.~Ichie, S.~Sasaki and H.~Toki, 
``Color Confinement and Hadrons", 
(World Scientific, 1995) 65.
\item{\rm 12.}
H.~Ichie, H.~Suganuma and H.~Toki, 
Phys.~Rev.~{\bf D52} (1995) 2944.
\item{\rm 13.}
H.~Suganuma, S.~Umisedo, S.~Sasaki, H.~Toki and O.~Miyamura,
Proc. of ``Quarks, Hadrons and Nuclei", 
Adelaide Australia, Nov. 1995, to appear in Aust. J. Phys. 
\item{\rm 14.}
H.~Ichie, H.~Monden, H.~Suganuma and H.~Toki, 
Proc. of ``Nuclear Reaction Dynamics of Nucleon-Hadron Many Body 
Dynamics", Osaka, Dec. 1995 (World Scientific) in press.
\item{\rm 15.}
N.~Seiberg and E.~Witten, Nucl.~Phys.~{\bf B426} (1994) 19; 
{\bf B431} (1994) 484. 
\nextline
G.~P.~Collins, Physics Today, March (1995) 17.
\item{\rm 16.}
Y.~Nambu, Phys.~Rev.~{\bf D10} (1974) 4262.
\nextline
G.~'t~Hooft, ``High Energy Physics" 
(Editorice Compositori, Bologna 1975).
\nextline
S.~Mandelstam, Phys.~Rep.~{\bf C23} (1976) 245.
\item{\rm 17.}
P.~A.~M.~Dirac, Proc.~Roy.~Soc.~{\bf A133} (1931) 60. 
\item{\rm 18.}
G.~'t~Hooft, Nucl.~Phys.~{\bf B190} (1981) 455.
\item{\rm 19.}
Z.~F.~Ezawa and A.~Iwazaki, Phys.~Rev.~{\bf D25} (1982) 2681; 
{\bf D26} (1982) 631.
\item{\rm 20.}
O.~Miyamura, Nucl.~Phys.~{\bf B}(Proc.~Suppl.){\bf 42} (1995) 538.
\nextline
O.~Miyamura and S.~Origuchi, {\it Color Confinement and Hadrons}, 
(World Scientific, 1995) 235.
\item{\rm 21.}
H.~Suganuma, A.~Tanaka, S.~Sasaki and O.~Miyamura, 
Proc. of ``Lattice Field Theories '95", 
Nucl.~Phys.~{\bf B}~(Proc.~Suppl.)~{\bf 47} (1996) 302.
\item{\rm 22.}
H.~Suganuma, K.~Itakura, H.~Toki and O.~Miyamura, 
Proc. of ``Non-perturbative Approaches to QCD", 
Trento Italy, July 1995 (PNPI, 1995) 224. 
\item{\rm 23.}
S.~Sasaki, H.~Suganuma and H.~Toki, Prog.~Theor.~Phys. {\bf 94} (1995) 
373.
\item{\rm 24.}
S.~Sasaki, H.~Suganuma and H.~Toki, Proc. of Int.~Conf. on 
``Baryons '95", Santa Fe, Oct. 1995 in press.
\item{\rm 25.}
H.~Ichie, H.~Suganuma and H.~Toki, preprint, 
Phys.~Rev.~{\bf D} in press. 
\item{\rm 26.}
T.~Kajino and R.~N.~Boyd, Ap.~J.~{\bf 359} (1990) 267.
\nextline
T.~Kajino, G.~J.~Mathew and G.~M.~Fuller, Ap.~J.~{\bf 364} (1990) 7.
\item{\rm 27.}
M.~Gao, Nucl.~Phys.~{\bf B9} (Proc. Suppl.) (1989) 368.
\item{\rm 28.}
Y.~Iwasaki, K.~Kanaya, L.~Karkkainen, Phys.~Rev.~{\bf D49} (1994) 3540.
\item{\rm 29.}
G.~M.~Fuller, G.~J.~Mathews, C.~R.~Alcock,
Phys.~Rev.~{\bf D 37} (1988) 1380.
\item{\rm 30.}
E.~Witten, Phys.~Rev.~{\bf D30}~(1984) 272. 
\item{\rm 31.}
T.~Kajino, Phys.~Rev.~Lett.~{\bf 66} (1991) 125.
\nextline
T.~Kajino, M.~Orito, Y.~Yamamoto and H.~Suganuma, 
``Color Confinement and Hadrons", (World Scientific, 1995) 263.

\endpage

\baselineskip15pt

\noindent
Table 1: Features of the lattice QCD and the QCD effective models.

\vskip0.5cm

\noindent
Table 2: The comparison among the ordinary electromagnetic system, 
the superconductor and the nonperturbative QCD vacuum 
regarded as the QCD-monopole condensed system. 

\vskip0.5cm

\noindent
Fig.1: A brief sketch of the QCD physics.

\vskip0.5cm

\noindent
Fig.2: A brief history of QCD and typical QCD effective models. 

\vskip0.5cm

\noindent
Fig.3 : The electric and magnetic inter-charge potentials 
in the Coulomb, Higgs and dual Higgs phases. 
The ``see-saw relation'' is found between in the electric and 
magnetic sectors. The long-range confinement system in one sector 
corresponds to a short-range (Yukawa-type) interaction system in the 
other sector.

\vskip0.5cm

\noindent
\hangindent=-7cm \hangafter=0
Fig.4: The string tension $k(T)$ at finite temperature $T$.
The black dots denote lattice QCD data [\gao] in the pure gauge. 

\vskip0.5cm

\noindent
\hangindent=-7cm \hangafter=0
Fig.5: The energy density ${\cal E}(r;R)$ of the hadron bubble.
It is negative inside and positive near the boundary surface.

\vskip0.5cm

\noindent
\hangindent=-7cm \hangafter=0
Fig.6: The energy of the hadron bubble $E(R;T)$ 
as function of the hadron bubble radius $R$.

\vskip0.5cm

\noindent
\hangindent=-7cm \hangafter=0
Fig.7: The critical radius $R_c(T)$, which provides 
maximum of $E(R;T)$, at finite temperature $T$. 

\vskip0.5cm

\noindent
\hangindent=-7cm \hangafter=0
Fig.8: The hadron-bubble formation factor 
$P(T) \equiv {\rm exp}(- {4 \pi \over 3} R_c^3(T) h(T)/T)$ 
as a function of temperature $T$.

\vskip0.5cm

\tenpoint

\noindent
Fig.9: The appearance and evolution of hadron bubbles 
in the early Universe. The shaded and white regions denote 
the hadron and QGP phases, respectively. 
(a) Slightly below $T_{\rm c}$, only large hadron bubbles 
appear with very small creation rate.
(b) Hadron bubbles expand with radiating shock wave, 
which reheats the QGP phase around them.
The affected region is expressed by the arrow.
(c) Near $T_{\rm low}$, many small hadron bubbles 
are violently created in the unaffected region. 
(d) The QGP phase pressured by the hadron phase 
is isolated as high-density QGP bubbles.

\end